\documentclass [hidelinks,12pt,openbib]{article}

\usepackage{cancel}
\usepackage{xcolor}
\usepackage[hidelinks]{hyperref}
\usepackage[normalem]{ulem}
 
\usepackage{mathtools}
 \usepackage{tikz}
 \usepackage{cancel}
\usepackage{fixmath}
\usepackage{dsfont}
\usepackage{caption}
\usepackage{pslatex}
\usepackage{bm}
\usepackage{graphicx}
\usepackage{amsmath,amssymb,amsfonts,amsthm}
\usepackage{xcolor}
\usepackage{mathtools}
\usepackage{dsfont}
\usepackage{cite}
\usepackage{ulem}
\usepackage{fancybox,fancyvrb}
\usepackage{framed} 

\usepackage{MnSymbol,wasysym}
\usepackage[export]{adjustbox}

\usepackage{enumitem}
\usepackage{color}
\usepackage{fancyhdr}
\usepackage{esint}
\usepackage{mathtools}
\usepackage[margin=1in]{geometry}
\usepackage{mdframed}

\def\XXint#1#2#3{{\setbox0=\hbox{$#1{#2#3}{\int}$}
     \vcenter{\hbox{$#2#3$}}\kern-.5\wd0}}
 \def\begfrwhite{\begin{mdframed}[
	backgroundcolor=white!15!white,
	linecolor=black,
	linewidth=0.5pt,
	align=center,
	userdefinedwidth=4.99in]}

\def\E1{\text{E}_1}

\def\L1loc{{L^1_{\rm loc}}}

\def\bc{\begin{center}}
\def\ec{\end{center}}

\def\begfr{\begin{mdframed}[
	backgroundcolor=gray!15!white,
	linecolor=blue,
	linewidth=1pt,
	align=center,
	userdefinedwidth=4.99in]}
\def\begfrwhite{\begin{mdframed}[
	backgroundcolor=white!15!white,
	linecolor=black,
	linewidth=0.5pt,
	align=center,
	userdefinedwidth=4.99in]}

\def\begfrblue{\begin{mdframed}[
	backgroundcolor=blue!15!white,
	linecolor=blue,
	linewidth=1pt,
	align=center,
	userdefinedwidth=4.99in]}

\def\endfr{\end{mdframed}}
\def\endfrred{\end{mdframed}}
\def\begfrred{\begin{mdframed}[
	backgroundcolor=red!15!white,
	linecolor=blue,
	linewidth=1pt,
	align=center,
	userdefinedwidth=4.99in]}
\def\endfrblue{\end{mdframed}}

\def\begeq{\begin{equation}}
\def\endeq{\end{equation}}

\def\bbr{\begin{bmatrix*}[r]}
\def\ebr{\end{bmatrix*}}
\def\bb{\begin{bmatrix*}[r]}
\def\eb{\end{bmatrix*}}
\def\bbc{\begin{bmatrix}}

\def\ebc{\end{bmatrix}}

\definecolor{darkmagenta}{rgb}{0.55, 0.0, 0.55}

\definecolor{darkgreen}{rgb}{0.0,0.6,0.0}

\def\R{\mathbb{R}}

\def\d{\displaystyle}

\def\0u{\underline{0}}

\def\begeq{\begin{equation}} 
\def\endeq{\end{equation}}

\def\bcenter{\begin{center}}
\def\ecenter{\end{center}}
\def\beq{\begin{equation}}
\def\eeq{\end{equation}}
\def\bmatr{\begin{bmatrix*}[r]}
\def\ematr{\end{bmatrix*}}
\def\bmatc{\begin{bmatrix}}
\def\ematc{\end{bmatrix}}

\usepackage{amsmath}               
  {
      \theoremstyle{plain}
      
  }
  {
      \theoremstyle{plain}
      
  }
   {
      \theoremstyle{plain}
      
  }
   {
      \theoremstyle{plain}
      
  }
  {
      \theoremstyle{plain}
      
  }
    {
      \theoremstyle{plain}
      
  }
\newcommand{\vertiii}[1]{{\left\vert\kern-0.25ex\left\vert\kern-0.25ex\left\vert #1 
    \right\vert\kern-0.25ex\right\vert\kern-0.25ex\right\vert}}
 \usepackage{amsthm}
\newtheorem*{proposition*}{Proposition}
\newtheorem*{corollary*}{Corollary}
\newtheorem*{definition*}{Definition}
\newtheorem*{example*}{Example}
\newtheorem*{lemma*}{Lemma}
\newtheorem*{theorem*}{Theorem}
\newtheorem*{observation*}{Observation}
\newtheorem*{remark*}{Remark}
\usepackage{sectsty}
\sectionfont{\fontsize{15}{10}\selectfont}

\newtheorem*{exmp*}{Example}
\usepackage{stmaryrd}
\usepackage{wrapfig}

\usepackage{xurl} 

\begin{document}

\begin{center}
{\bf \Large A new probabilistic analysis of the yard-sale model } 
\vskip 15pt
Christoph B\"orgers$^1$ and Claude Greengard$^{2}$
\end{center}

\noindent
$^1$ Department of Mathematics, Tufts University, Medford, MA 

\noindent
$^2$ Two Sigma Investments, LP, New York, NY, and 

\noindent
\hskip 10pt Courant Institute of Mathematical Sciences, New York University, New York, NY

\vskip 20pt

\begin{quote}
{\small {\bf Abstract.} } In Chakraborti's {\em yard sale model} of an economy \cite{original_yard_sale}, identical agents engage
 in trades that result in wealth exchanges,
but conserve  the combined wealth of all agents and each agent's {\em expected} wealth. In this model, 
{\em wealth condensation}, that is, convergence to a state in which one agent owns everything and the others own
nothing, occurs almost surely. We give a proof of this fact that is much shorter
than existing ones and extends to a modified model in which there is a wealth-acquired
advantage, i.e., the wealthier of two trading partners is more likely to benefit from the trade.
\end{quote}

\vskip 20pt
\section{Background}

The yard-sale model, first proposed in  \cite{original_yard_sale}, is a caricature 
of a set of agents trading with each other. The agents are identical, and there are $N$ of them. 
Time proceeds in discrete steps labeled $0,1,2,\ldots$, and in the $n$-th step the $i$-th agent has wealth $X_n^i \geq 0$. 
We will write 
$$
X_n = \left[ X_n^i \right]_{1 \leq i \leq N} \in \R^N.
$$
Without loss of
generality, we assume 
$$
\sum_{i=1}^N X_0^i = 1.
$$
We think of $X_0$ as deterministic. The $X_n$ with $n \geq 1$ are random, defined inductively as follows.
Given $X_{n-1}$, choose a random pair of integers, $(\mu_n, \nu_n)$, uniformly distributed 
in the set of all pairs $(i,j)$ with $1 \leq i,j \leq N$,  $i \neq j$. 
Without loss of generality, assume $X_{n-1}^{\mu_n} \leq X_{n-1}^{\nu_n}$. 
So agent $\nu_n$ is, at time $n-1$, at least as wealthy as agent $\mu_n$.
Imagine that agents $\mu_n$ and $\nu_n$ now engage in an economic transaction.
As a result of errors (perhaps over- or underpayments occurring because of lack of 
complete information for instance),  a random fraction $B_n \in (0,1)$ of the wealth of agent $\mu_n$ (the poorer of the two) is transferred either 
from agent $\mu_n$ to agent $\nu_n$, or vice versa. The $B_n$ are assumed to be identically distributed.
The direction of the transfer is determined by a fair coin flip. 
 Formally, let $V_n=1$ with probability $1/2$, and $V_n=-1$ otherwise. Then
$$
X_n^{\mu_n} = X_{n-1}^{\mu_n} +V_n ~\! B_n ~\!  X_{n-1}^{\mu_n} ~~~~~~\mbox{and} ~~~~~~
X_n^{\nu_n} = X_{n-1}^{\nu_n} -V_n ~\! B_n ~\!  X_{n-1}^{\mu_n} .
$$
The pairs $(\mu_n, \nu_n)$, the fractions $B_n$, and the signs $V_n$ are assumed to be
independent of each other and of the $X_k$, $B_k$, and $V_k$ with $k \leq n-1$.
In reference \cite{original_yard_sale}, and in much of the literature on the yard-sale model, $B_n$ is assumed to be a 
deterministic number $\beta \in (0,1)$. We use a random $B_n \in (0,1)$ more generally. Notice that wealth is conserved:
$$
\sum_{i=1}^N X_n^i = 1
$$
for all $n$. 
The model is known to have the following surprising property, which we will call the {\em yard-sale theorem}.

\vskip 5pt
\noindent
{\bf Yard-Sale Theorem.} {\em 
{\rm (a)}  There almost surely exists an $i \in \{1,\ldots,N\}$ with 
$\d{
 \lim_{n \rightarrow \infty} X_n^i = 1 .}
$

\noindent
{\rm (b)} For all $i \in \{1,\ldots,N\}$, 
\begin{equation}
\label{eq:who_wins}
P \left( \lim_{n \rightarrow \infty} X_n^i = 1 \right) = X_0^i. 
\end{equation}
}

In the limit, one agent owns everything. This sort of maximal inequality is called {\em wealth condensation} in the literature. 
In the yard-sale model, wealth condensation is the inescapable result of random, statistically unbiased interactions.

For a version of the model in which there is a {\em continuum} of agents, rather than a finite number, a result of this kind 
was proved by Boghosian {\em et al} \cite{Boghosian_Johnson_Marcq}. Chorro 
 \cite{chorro} pointed out that the theorem as stated above  is an immediate consequence of Doob's martingale convergence theorem:
For a fixed $i$, the sequence $\{X_n^i \}_{n=0,1,2,\ldots}$ is a bounded martingale, and therefore must converge. It is
clear from the definition of the model that the $X_n^i$ cannot all converge unless one converges to 1 and the others converge to 0. Equation (\ref{eq:who_wins}) follows
from the fact that $E(X_n^i) = E(X_0^i)$ for all $n$ and $i$.

Some interesting variations can be handled  immediately using the same reasoning. For instance, different agents can be assumed
to have different degrees of
risk tolerance  \cite{cardoso_et_al_2023}. The amount of wealth transferred during the trade between agents $\mu_n$ and $\nu_n$ might be taken to be 
$V_n ~\! B_n ~\! \lambda_{\mu_n} ~\! X_{n-1}^{\mu_n}$, where the $\lambda_i$, $1 \leq i \leq N$, are fixed numbers in $(0,1)$; 
if $\lambda_i$ is smaller, agent $i$ is more risk averse.
Obviously but remarkably, eq.\ (\ref{eq:who_wins}) still holds for the modified model; risk aversion
does not make an agent less or more likely to end up owning everything.

Cardoso {\em et al.} \cite{cardoso_et_al_2023} have recently proposed a different argument, based on the Gini index, to derive related results for
a broader class of models. Their analysis relies on what they call the  {\em fair rule hypothesis} \cite[Equation (8)]{cardoso_et_al_2023}. 
For wealth-conserving models, it is the martingale property.  
Most, but not all, examples in \cite{cardoso_et_al_2023} are wealth-conserving. The result in \cite{cardoso_et_al_2023} is that the Gini index is monotonically
increasing, and stationary if and only if it is 1.

An interesting extension is obtained by adding a {\em wealth-acquired advantage}  \cite{Boghosian_and_gang}.
The coin flips that determine in which direction the wealth will flow in each interaction is biased
in favor of the wealthier agent: 
$V_n = 1$ with probability $p$, and $V_n = -1$ otherwise, where $p$ is no longer required to be $\frac{1}{2}$, but is allowed to be anywhere in $\left[ \frac{1}{2}, 1 \right)$. 
One should certainly expect wealth condensation for $p>\frac{1}{2}$ if there is wealth condensation 
for $p = \frac{1}{2}$. However, proofs that rely on the martingale property no longer work; the model is still wealth-conserving, but $\{X_n^i\}_{n \geq 0}$ is no longer
a martingale, nor a sub- or super-martingale.

To the model with a (possible) wealth-acquired advantage, 
Boghosian {\em et al.}\  \cite{Boghosian_2014,Boghosian_2014a} added {\em wealth taxation}. 
Each agent is taxed
a fraction $\chi \in (0,1)$ of their fortune in each time step, uniformly re-distributing the total amount taken in: 

\pagebreak

$$
\tilde{X}_n^{\mu_n} = X_{n-1}^{\mu_n} +V_n ~\! B_n~\! X_{n-1}^{\mu_n} ~~~~~~\mbox{and} ~~~~~~
\tilde{X}_n^{\nu_n} = X_{n-1}^{\nu_n} -V_n ~\! B_n~\! X_{n-1}^{\mu_n}, 
$$
$$
X_n^i = (1-\chi) \tilde{X}_n^i+ \frac{\chi}{N} ~~~\mbox{for all $i \in \{1,\ldots,N\}$}.
$$
It  is clear that wealth taxation, 
 no matter how small $\chi \in (0,1)$ may be, prevents wealth condensation: We now have 
$$
X_n^i \geq \frac{\chi}{N}
$$
for all $n \geq 1$ and $i \in \{1,\ldots,N\}$, which  precludes the existence of the limits $\lim_{n \rightarrow \infty} X_n^i$.

For a version of the model with wealth-acquired advantage and taxation in which there is a continuum of agents, 
Boghosian {\em et al} \cite{Boghosian_and_gang} have shown that for each $p \in \left( \frac{1}{2}, 1 \right)$, there is a threshold value $\chi_c$ depending on $p$, 
such that there will be {\em oligarchy} for $\chi < \chi_c$, but not for $\chi>\chi_c$. Here {\em oligarchy} means that 
a vanishingly small fraction of the population will own, in the long run, 
a non-vanishing fraction of total wealth.

In this brief note, we propose a new probabilistic proof of part (a) of the Yard-Sale Theorem, which applies also if $p \in \left( \frac{1}{2}, 1 \right)$. 
Our proof does not use the martingale property, and therefore applies when $p \in \left( \frac{1}{2}, 1 \right)$. The tools used in our analysis are much 
lighter than those used in previously published proofs of the Yard-Sale Theorem or similar results.  Instead of the Gini index, 
we use $\| X_n \|^2$ as the measure of concentration, where $\| \cdot \|$ denotes the Euclidean norm.  
We note that the idea of using the Euclidean norm of a probability vector as a measure of concentration is not new; it appears  in 
quantum physics \cite{inverse_participation_ratio}, political science \cite{Laakso_Taagepera}, ecology \cite{simpson_diversity}, and antitrust regulation \cite{justice_department}.

\section{Proof of almost sure wealth condensation} 
\label{sec:alternative}

We consider, from here on, the yard-sale model, possibly with a wealth-acquired advantage: $\frac{1}{2} \leq p < 1$, but without taxation.
Write $M_n = \max_i X_n^i$. We will prove $M_n \rightarrow 1$ almost surely, which is part (a) of the Yard-Sale Theorem. 
We have no analogue of part (b) for $p>\frac{1}{2}$.

We begin with a preliminary calculation in which we determine  the expected change in the concentration measure $\| X_n \|^2$ in a single step. 
(As before, $\| \cdot \| $ is the Euclidean norm.) 
Suppose that $X_{n-1}^i$, $1 \leq i \leq N$, are given,  $\mu_n$ and $\nu_n$ have been chosen, with 
$X_{n-1}^{\mu_n} \leq X_{n-1}^{\nu_n}$, and $B_n$ has been chosen as well. We write $\omega_n = B_n ~\! X_{n-1}^{\mu_n}$; this is the amount of wealth
at stake in the trade between agents $\mu_n$ and $\nu_n$. We also write $p = \frac{1}{2} + \delta$, with $0 \leq \delta \leq \frac{1}{2}$. 
Then the (conditional) expectation of $\| X_n\|^2 - \| X_{n-1}\|^2$ equals
$$
\left( \frac{1}{2} + \delta \right)  \left( \left( X_{n-1}^{\mu_n} -\omega_n \right)^2 - \left( X_{n-1}^{\mu_n} \right)^2  + 
\left(X_{n-1}^{\nu_n} + \omega_n \right)^2 - \left( X_{n-1}^{\nu_n} \right)^2 \right) + 
$$
$$
\left( \frac{1}{2} - \delta \right) \left( \left( X_{n-1}^{\mu_n} +\omega_n \right)^2 - \left( X_{n-1}^{\mu_n} \right)^2  + 
\left(X_{n-1}^{\nu_n} - \omega_n \right)^2 - \left( X_{n-1}^{\nu_n} \right)^2 \right) = 
$$
$$ 
2 \omega_n^2 + 4 \delta \omega_n \left(X_{n-1}^{\nu_n} - X_{n-1}^{\mu_n} \right) \geq 2 \omega_n^2.
$$
This implies 
\begin{equation}
\label{eq:bound_on_increase}
E \left( \| X_n \|^2  \right) -  E \left( \| X_{n-1} \|^2 \right) \geq 2 E(\omega_n^2)
\end{equation}
where $E$ denotes {\em unconditional} expectations.

We now proceed to the main argument, which makes use of inequality (\ref{eq:bound_on_increase}) at the end.  To show $M_n \rightarrow 1$ almost surely, 
it is enough to show that $\omega_n \rightarrow 0$ almost surely. 
Therefore we have to show that for any $\epsilon>0$, it almost surely happens only finitely many times that 
$\omega_n \geq \epsilon$.
By the Borel-Cantelli lemma, it is  enough to show 
$$
\sum_{n=1}^\infty P(\omega_n \geq \epsilon) <\infty
$$
for any $\epsilon>0$. Since 
$
E(\omega_n^2) \geq P(\omega_n \geq \epsilon) \epsilon^2,
$
we have 
$$
\sum_{n=1}^\infty P(\omega_n \geq \epsilon) \leq \frac{1}{\epsilon^2} \sum_{n=1}^\infty E(\omega_n^2). 
$$
Using (\ref{eq:bound_on_increase}), 
$$
\sum_{n=1}^\infty E(\omega_n^2) \leq \frac{1}{2}  \sum_{n=1}^\infty  \left( E(\| X_n \|^2 - E(\| X_{n-1} \|^2) \right)  \leq 
\frac{1}{2} \left( \limsup_{n \rightarrow \infty} E(\| X_n \|^2) - E(\| X_0 \|^2)  \right) \leq  \frac{1}{2}.
$$
This  completes the proof.

%\bibliography{/Users/cborgers/Documents/refs.bib}

\begin{thebibliography}{10}

\bibitem{Boghosian_2014a}
B.~Boghosian.
\newblock {Fokker-Planck description of wealth dynamics and the origin of
  Pareto's law}.
\newblock {\em Int. J. Mod. Phys. C}, 25(12):1441008, 2014.

\bibitem{Boghosian_2014}
B.~Boghosian.
\newblock Kinetics of wealth and the {P}areto law.
\newblock {\em Phs. Rev. E}, 89, 042804, 2014.

\bibitem{Boghosian_Johnson_Marcq}
B.~Boghosian, M.~Johnson, and J.~Marcq.
\newblock {An H theorem for Boltzmann's equation for the yard-sale model of
  asset exchange}.
\newblock {\em J. Stat. Phys.}, 161:1339--1350, 2015.

\bibitem{Boghosian_and_gang}
B.~M. Boghosian, A.~Devitt-Lee, M.~Johnson, J.~A. Marcq, and H.~Wang.
\newblock {Oligarchy as a phase transition: The effect of wealth-attained
  advantage in a Fokker-Planck description of asset exchange}.
\newblock {\em Physica A}, 476:15--37, 2017.

\bibitem{cardoso_et_al_2023}
B.-H.~F. Cardoso, S.~Gon\c{c}alves, and J.~R. Iglesias.
\newblock {Why equal opportunities lead to maximum inequality? The wealth
  condensation paradox generally solved}.
\newblock {\em Chaos Solit. Fractals}, 168, 113181, 2023.

\bibitem{original_yard_sale}
A.~Chakraborti.
\newblock {Distributions of money in model markets of economy}.
\newblock {\em Int. J. Mod. Phs. C}, 13(10):1315--1321, 2002.

\bibitem{chorro}
C.~Chorro.
\newblock A simple probabilistic approach of the yard-sale model.
\newblock {\em Stat.\ Probab.\ Lett.}, 112:35--40, 2016.

\bibitem{inverse_participation_ratio}
B.~Kramer and A.~MacKinnon.
\newblock {Localization: theory and experiment}.
\newblock {\em Rep. Prog. Phys.}, 56:1469--1564, 1993.

\bibitem{Laakso_Taagepera}
M.~Laakso and R.~Taagepera.
\newblock {``Effective" number of parties: a measure with applications to West
  Europe}.
\newblock {\em Comp. Polit. Stud.}, 12(1):3--27, 1979.

\bibitem{simpson_diversity}
E.~H. Simpson.
\newblock {Measurement of diversity}.
\newblock {\em Nature}, 163, 1949.

\bibitem{justice_department}
{United States Department of Justice}.
\newblock {Herfindahl-Hirschmann Index}.
\newblock https://www.justice.gov/atr/herfindahl-hirschman-index.

\end{thebibliography}
%\bibliographystyle{plain}

 \end{document}